\documentclass[twocolumn,prb,multicol,amsmath,amssymb]{revtex4}
\usepackage[dvips]{graphicx}
\usepackage{graphicx}
\usepackage{dcolumn}
\usepackage{bm}
\usepackage{graphics}
\usepackage{epsfig,color}

\newcommand{\be}{\begin{equation}}
\newcommand{\ee}{\end{equation}}
\newcommand{\bea}{\begin{eqnarray}}
\newcommand{\eea}{\end{eqnarray}}

\newcommand{\bS}{\bf S}

\begin{document}

\title[J. Vahedi ]{Quantum Chaos in the Heisenberg Spin Chain: the Effect of Dzyaloshinskii-Moriya Interaction}
\author{J. Vahedi$^{1}\footnote{email: javahedi@gmail.com\\ Tel: (+98)9111554504\\Fax: (+98)151 33251506}$, A. Ashouri$^{2}$ and S. Mahdavifar$^{2}$}
\address{$^{1}$Department of Physics, Sari Branch, Islamic Azad University, Sari, Iran.}
\address{$^{2}$Department of Physics, University of Guilan , 45196-313, Rasht\emph{}, Iran}
\date{\today}
\begin{abstract}
Using one-dimensional spin-1/2 systems as prototypes of quantum many-body systems, we study the emergence of quantum chaos. The main purpose of this work is to answer the following question: how does the spin-orbit interaction, as a pure quantum interaction, may lead to the onset of quantum chaos? We consider three integrable spin-1/2 systems: the Ising, the XX, and the XXZ limits, and analyze whether quantum chaos develops or not after the addition of the Dzyaloshinskii-Moriya interaction. We find that, depending on the strength of the anisotropy parameter, the answer is positive for the XXZ and Ising models, while no such evidence is observed for the XX model. We also discuss the relationship between quantum chaos and thermalization.
\\
\\
 \textbf{The main purpose of this work is to answer this question: how the spin-orbit interaction as a pure quantum interaction may develop a quantum chaos which has no classical counterpart?. The result can be summarized as follows:
\\1- Ising chain with added Dzyaloshinskii-Moriya (DM) Interaction is chaotic.
\\2- XX chain with added DM interaction does not show a chaotic features.
\\3- XXZ chain with added DM interaction is chaotic.}
\end{abstract}
\maketitle

\section{Introduction}\label{sec1}
Quantum chaos began as an attempt to find chaos, in the sense of extreme sensitivity to changes in initial conditions, in quantum mechanical systems. The most natural guess would be that chaos would appear in the quantum realm in much the same way as it appears in the classical one, namely in the form of sensitive dependence on initial conditions.  However, the laws of quantum mechanics do not permit a similar definition, since the laws do not show an exponential sensitivity to the initial conditions. So it boosts this question, how does classical chaos emerge from a quantum system? Are there any signatures of chaos in a quantum system whose classical counterpart is chaotic?\cite{a1,a1-1,a1-2,a1-3}
\par
Over recent decades, researchers have gained some important insights about the relationship between quantum mechanics and classical chaos. One of the most solid conclusions concern the statistical properties of the energy levels of large quantum systems\cite{a2}, such as heavy atomic nuclei\cite{a2-0,a2-00}. If a classical non-chaotic system is quantized, the resulting discrete energy values (eigenvalues) tend to cluster, while eigenvalues in quantum systems whose classical counterparts are chaotic typically repel each other and tend to be distributed more uniformly within the available energy range.
\par
The general integrable systems have the Poisson distribution of energy level spacing, which implies the existence of many degenerate eigenenergies. On the other hand, the quantum chaotic systems have the Wigner-Dyson distribution of energy level spacing. The Wigner-Dyson distribution has two main features:  zero probability at zero energy level spacing and a peak at a finite energy level spacing. The former feature means that there is a little degeneracy while the latter implies that there are a large number of energy level spacings around the peak value. These signatures of chaos have been experimentally observed in the energy levels obtained from molecular and atomic spectra\cite{a2-3,a2-4}.
 \par
 Low-dimensional many body quantum systems,  specially, spin-1/2 chains, are very good candidates\cite{Hsu93, Emary03, Zangara13, Santos04, Avishai02, Kudo05, Boness07, Znidaric08, Santos10, Rigol10} for  studying quantum chaos. It is known that the level spacing distribution for the integrable XXZ chain is Poissonian\cite{Hsu93} for all values of the anisotropy parameter, apart from some special ones that may lead to excessive degeneracies\cite{Zangara13}. In 2004, Santos showed that a transition to the chaotic phase happens with the addition of a single defect in the spin-1/2 XXZ chain model\cite{Santos04}. One should note that, an open chain with defects only on edges is not chaotic.  In addition, studies show that  the transition to chaos can also be achieved with onsite disorder \cite{Avishai02}, by  increasing the strength of the next-nearest neighbor interaction\cite{Hsu93,Santos10, Rigol10}, or by coupling a second chain\cite{Hsu93}. 
 \par 
On the other hand,  motivated by some recent experimental realization of spin-orbit coupled optical lattices for both fermions\cite{Wang12,Cheuk12} and bosons\cite{Jimenez12}, author in ref[\cite{Gong15}] synthesize spin-orbit and Zeeman coupling into an effective Hamiltonian for bosons and spin-1/2 fermions in optical lattices. It is shown that spin-orbit coupling leads to an effective in-plane  Dzyaloshinskii-Moriya (DM) term, which its strength predicted of the same order as the Heisenberg coupling constant. Moreover, with the advent of experiments on optical lattices (which mimic the lattices in real materials), can enable to explore condensed matter phenomena in a controlled and almost pure environment. The general hope is to be able to simulate interesting many-body systems beyond the reach of state-of-the art numerical methods and to find answers to so far unsolved questions of the field.
 \par
In this work, we explore whether the onset of chaos can also take place in spin-1/2 systems  if one includes also the  DM interaction. From a physical point of view, this interaction is one of the agents responsible for magnetic frustration. Since this interaction may induce spiral spin arrangements in the ground state, it is closely involved with ferroelectricity in multiferroic spin chains\cite{Seki08, Huvonen09}. Moreover, the DM interaction modifies the dynamic properties and the quantum entanglement of spin chains\cite{Derzhko06, Kargarian09, Vahedi12, Soltani14}.  Ising model with the DM interaction was extensively studied\cite{Soltani12}.
\par
We consider the three limits of the 1D Heisenberg model, namely the Ising, XX, and XXZ models. They have been previously explored before in the context of quantum chaos and quantum information \cite{a2-5,a2-6,a2-7, a2-8,a2-9,a2-10}. We analyze the effects of the DM interaction on these models with focus on the level spacing statistics and localization in real space. Using exact diagonalization, we find that the Ising and XXZ models can develop chaos, but the same does not happen for the XX model. In terms of localization we find the structures of the eigenstates in real space are very similar when the system is chaotic, but differ in the integrable limit. We also address the relationship between chaos and quantum thermalization in isolated interacting many-body systems based on eigenstate thermalization hypothesis(ETH). It is found that the onset of chaos is intimately related with the viability of thermalization and this indicates that the ETH should be valid in the chaotic domain.
 \par
The paper is organized as follows. In section (II), we summarize the model Hamiltonian. In section (III), the approach and formalism are addressed. In section (IV), we present our numerical results. Finally, the conclusion is given in section (V).
\section{THE MODEL}\label{sec2}
We consider a spin-1/2 chain with only nearest neighbor interaction. The Hamiltonian of the anisotropic XXZ model in a uniform magnetic field (\emph{h}) with added the DM interaction and an open boundary chain of length \emph{L}, is given by
\begin{eqnarray}
{\cal H} =\sum_{n=1}^{L-1} J\Big({\bS}^{x}_{n}
{\bS}^{x}_{n+1}+{\bS}^{y}_{n}
{\bS}^{y}_{n+1}\Big)+J_{z}(S^{z}_{n}S^{z}_{n+1})\nonumber\\
+D \sum_{n=1}^{L-1}\Big({\bS}^{x}_{n}{\bS}^{y}_{n+1}
-{\bS}^{y}_{n}{\bS}^{x}_{n+1}\Big)
 + h\sum_{n=1}^{L} S^{z}_{n}+\epsilon_{d} S^{z}_{d}
\label{Hamiltonian}
\end{eqnarray}
where $S_{n}^{x,y,z}$ are spin-1/2 operators on the n-th site. $J>0$ and $J_{z}>0$ are the antiferromagnetic (AF) coupling constants. $J$ is the flip-flop coupling and responsible for propagating the excitation through the chain. A spin pointing up corresponds to an excitation. $J_{z}$ is the Ising coupling. We work with an anisotropic chain, that is, the coupling constant $J$ for the $XX$ type interaction is different to the coupling constant $J_{z}$ for diagonal Ising interaction. The term $h$ gives the Zeeman splitting of each spin $n$ subjected to a magnetic field in the $z$ direction. A defect corresponds to the site which has a larger Zeeman splitting $(h_{d}=h+\epsilon_{d}$) an it is caused by a magnetic field slightly larger than the field applied on the other sites. In the case $\epsilon=0$, the chain is ideal because all sites have the same energy splitting. 
\par
It is also worth to address some characteristics of the Ising and XXZ models compare with the XX model, regarding to the DM interaction. Using the raising and lowering operators, $S^+_n=S^+_n+iS^+_n$ and $S^-_n=S^+_n-iS^+_n$, one can rewrite the interaction part of the Eq.(\ref{Hamiltonian}) as $ \big(\frac{J+iD}{2}\big){\bS}^{x}_{n}{\bS}^{x}_{n+1}+\big(\frac{J-iD}{2}\big){\bS}^{y}_{n}{\bS}^{y}_{n+1}+J_{z}{\bS}^{z}_{n}{\bS}^{z}_{n+1}$. It is clear that the DM interaction adds some anisotropic features to the system due to the absences of inversion symmetry. So, the $XXZ$ model changes to the $XYZ$ model. Nevertheless, the diagonal Ising interaction term which favors localization exists without any modification and just the off-diagonal terms modified which favors to propagate excitation and consequently induce delocalization.
\par
 In this model, Eq.~(\ref{Hamiltonian}), the total spin in the $z$ direction, $S^{z}=\sum_{i=1}^{N}S^{z}_{i}$, is conserved: $[H,S^{z}]=0$, so states with different $S^{z}$ are not coupled. To obtain the level spacing distribution, we focus on a particular subspace of the Hamiltonian with size $L=15$, namely subspaces with $10$ spins up and $3003$ eigenvalues. 
\par
Theoretically, in the simplest case of the XXZ spin chain with nearest-neighbor interaction and in the absence of the defect, $\epsilon=0$, and the DM interaction, $D=0$, the model is integrable and it can be analytically solved using the Bethe ansatz\cite{a6, a7}. We also recall the behavior of the model in the absence of the DM interaction, $(D=0)$, but with the presence of a defect, $(\epsilon\neq0)$, reported by Gubin et al.\cite{a8} and for an isotropic chain by Santos\cite{Santos04}. Depending on the conditions bellow, the system defined by Eq.(\ref{Hamiltonian}) in the absence of the DM interaction can be chaotic:
\begin{itemize}
  \item The strength of the Ising interaction cannot be much larger than XY-type coupling
  \item The defect cannot be placed on the edges of the chain\cite{a9}
  \item  We cannot have defect energy much larger than flip-flop coupling.
 \end{itemize} 
The source of quantum chaos is the competition between the Ising interaction, the off-diagonal terms and the defect\cite{a10}.
\section{Approach}\label{sec2}
To discriminate between the chaotic and the integrable behavior of the system under consideration, one should work with some tools. Below, we give a concise explanation about our toolkit to deal with this problem.
\subsection{Level spacing distribution}
The first choice is the level spacing statistics which we have addressed in the introduction. But one notes that some extra manipulations are needed to harness the level spacing as a good tool to quantify the crossover from integrability to quantum chaos.  To find universal statistics within each invariant subspace, it is necessary to to unfold the spectrum, which consists of locally rescaling the eigenvalues $E_{i}$ so that the mean level density of the unfolded eigenvalues is equal to unity. There are different standard numerical unfolding procedures. One of these procedures is explained  with details in Ref.~[\onlinecite{a8}].
\par
 We consider the nearest-neighbor spacing distribution (NNSD) $P(s)$, where $s_{i}=\varepsilon_{i+1}-\varepsilon_{i}$ is the spacing between consecutive unfolded eigenvalues. For a quantum integrable system whose energy levels are not correlated, the distribution is typically Poissonian , $P_{poi}(s)=\exp(-s)$, while for chaotic quantum system whose energy levels are correlated and crossing are avoided, the level spacing distribution is characterized by the Wigner-Dyson surmise, $P(s)=(\pi s/2) \exp(-\pi s^{2}/4) $, the Wigner surmise for ensembles with time reversal invariance (GOE).
\subsection{Number of principal components}
Our second choice is the number of principal components (NPC). Contrary to the previous tool, the NPC  exploits the eigenstates statistics. For eigenstate $|\psi_i\rangle=\sum_{n=1}^{dim}c_{in}|\phi_n\rangle$, the NPC is defined as
\begin{equation}
NPC=\frac{1}{\sum_{n=1}^{dim}|c_{in}|^4}
\end{equation}
where $|\phi_n\rangle$ is the basis vectors. The NPC is a tool to measure the spreadness of eigenstates in the chosen basis. Indeed, this quantity gives the numbers of basis vectors that contribute to each eigenstate. It is small when the state is localized localized, in which case there are a few number of non-zero components c{in}. It is large when the state is delocalized, in which case a  large portion of the components are non-zero. Complete delocalization occurs for Gaussian Orthogonal Ensembles (GOE), where the eigenstates are random vectors, that is, the coefficients $c_{in}$ are independent random variables from a Gaussian distribution, the average over the ensemble leads to the number of principal components, $\overline{NPC}^{GOE}$, $\thicksim D/3$\cite{a10,a11}. 
\par
Note, the choice of basis is essential for studying the structure of the eigenstates, so we use the configuration space basis, which is also known as the site-basis\cite{Torres15}.  The site-basis vectors $|\phi_n\rangle$ correspond to states where the spin on each sith either pionts down or up along the $z$-axis as $|\downarrow\uparrow\downarrow\uparrow\dots>$.
\subsection{Thermalization}
In order to check the relationship between quantum chaos and quantum thermalization, we study the eigenstate thermalization hypothesis (ETH). Several theoretical works have focused on  the question of thermalization in nonintegrable quantum systems\cite{Rigol1-09, Rigol2-09, Gogolin16}. It is known that the onset of thermalization in isolated interacting many-body systems is related to a chaotic structure of many-body eigenstates in the selected basis. According to ETH, expectation values of physical observables in energy eigenstates are approximately smooth functions of their energy in the chaotic phase. Here we consider the magnetization on defect site as an observable which is defined as
\begin{equation}
\langle S_{d}^{z}\rangle=\langle \psi_i | S_{d}^{z} | \psi_i \rangle,
\end{equation}
where $\langle \psi_ \rangle$ is the eigenstate corresponds to the energy eigenvalue, $E_{i}$, of the system.
 \begin{figure}[h!]
\includegraphics[width=1.0\columnwidth]{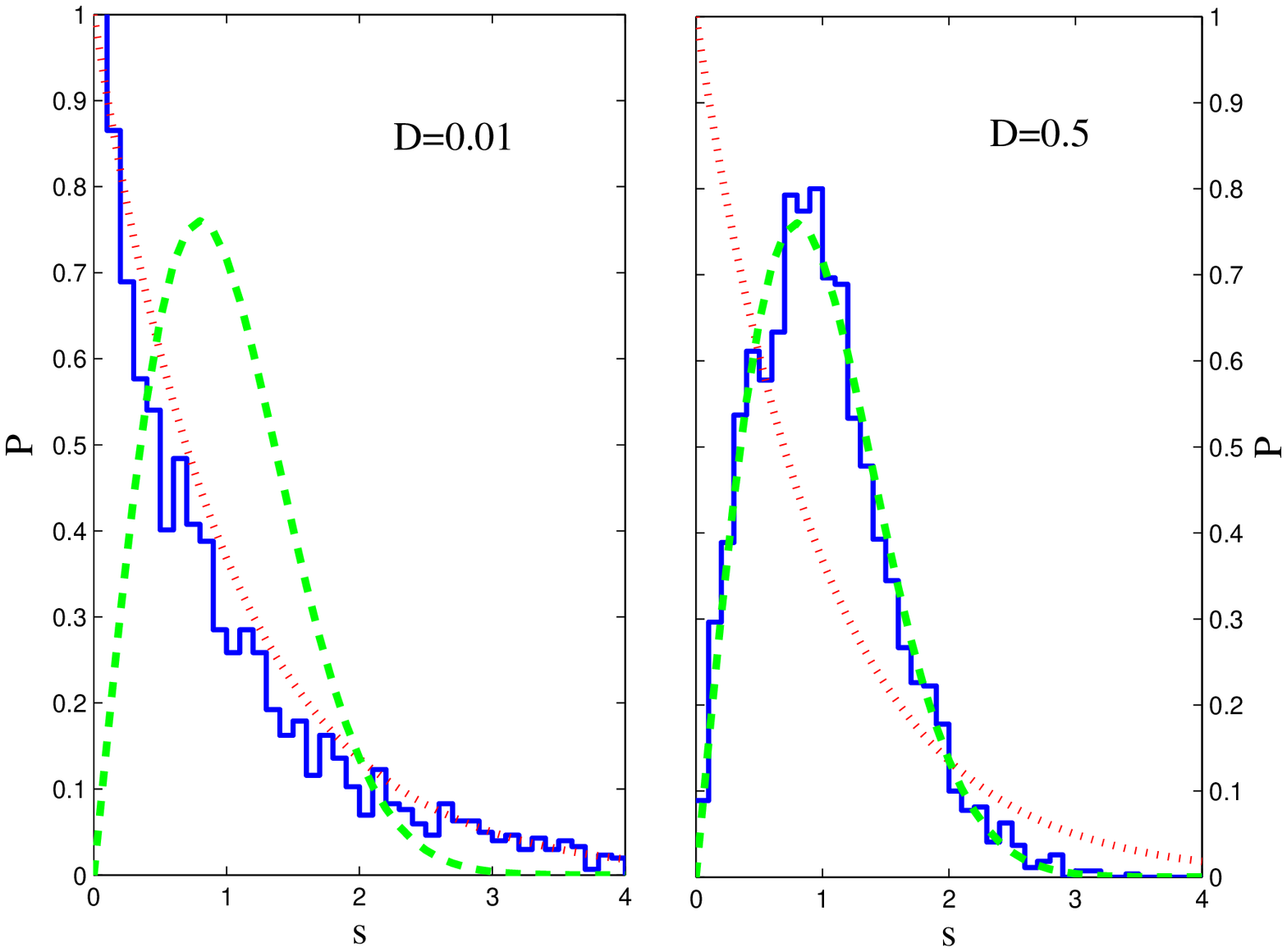}
\includegraphics[width=1.0\columnwidth]{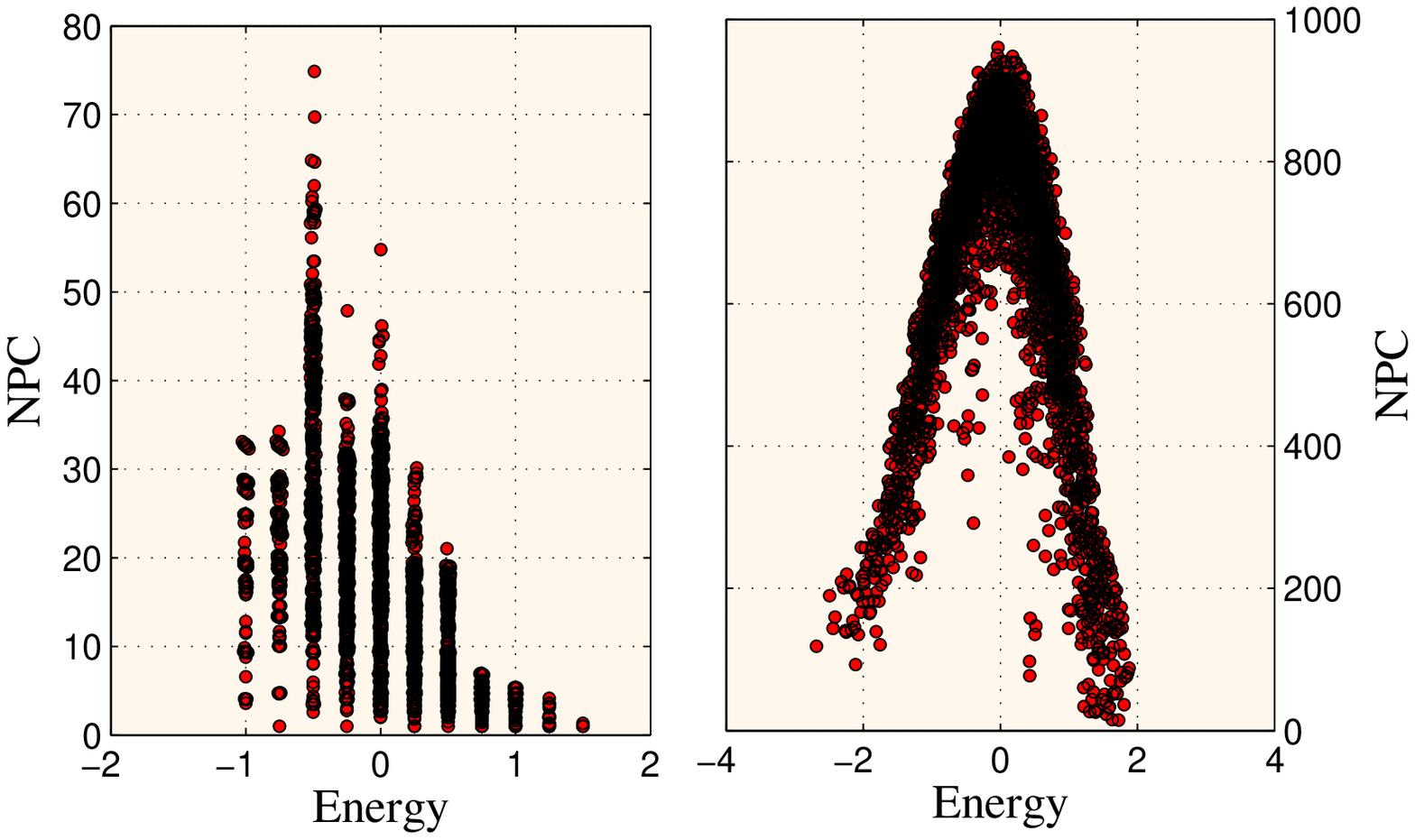}
\includegraphics[width=1.0\columnwidth]{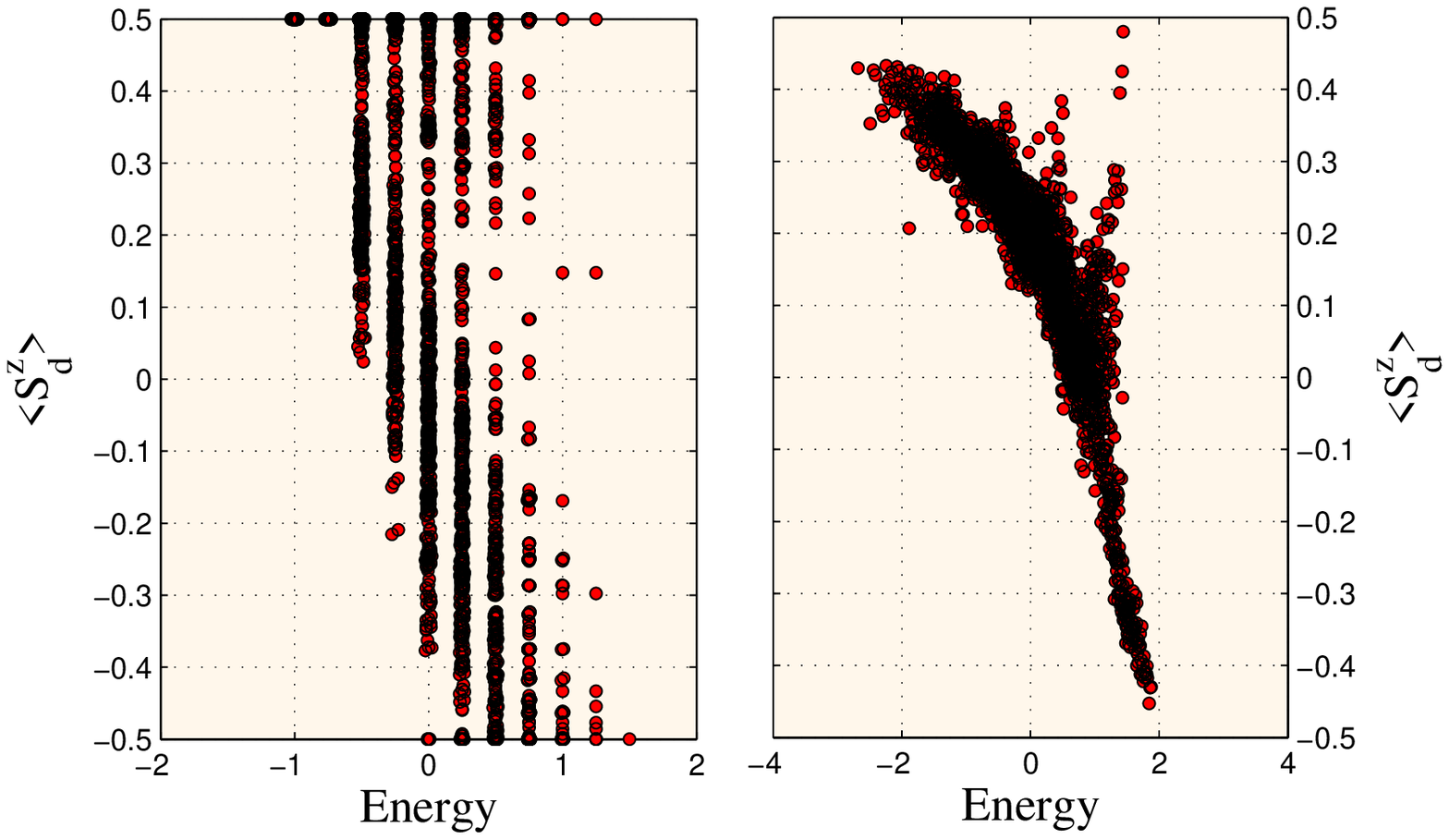}
\caption{(Color online) Top, middle and bottom panels  depict the level spacing distribution $P(s)$, the number of principle components NPC and $EEVs$ of $S_{d}^{z}$ vs energy for the full spectrum, respectively, for the Hamiltonian in Eq.~(\ref{Hamiltonian}) with $L=15$, $10$ up-spins up and for two values of  DM interaction (left column $D=0. 01$ and right column $D=0. 5$). Other parameters are set to $J=0$, $J_{Z}=0.5$, $\epsilon=0.5$, $bin size=0.1$,  and a defect on site $d=7$. In both level spacing distribution functions, (top panels), dotted line gives the Poisson distribution and long-dashed line corresponds to the Wigner-Dyson distribution.}
\label{fig1}
\end{figure}
\section{Numerical results}
 In this section, we present the numerical results obtained from the simulation using the exact diagonalization method. The mutual effect of a uniform magnetic field and the DM interaction on  the spacing distribution, the NPC of the system and the expectation value of the magnetization of the defect  are investigated.
\par
 \subsection{Ising model with the DM interaction}
First, we analyze the effect of the DM interaction on the Ising chain, so J=0, in the presence of an external magnetic field the $z$ direction and in the presence of a defect on site $d=[L/2]$. When $D=0$, the system is integrable and the spacing distribution is Poisson, as previously reported. The main question we are interested in is, how the DM interaction may affect on the system from the quantum chaos point of view.
 \par
 In Fig.~\ref{fig1}, we plot the level spacing distribution $P(s)$, the NPC distribution and the magnetization on the defect site, $\langle S_{d}^{z}\rangle=\langle \psi_i | S_{d}^{z} | \psi_i \rangle$,  for chain with length $L=15$  and for different values of the DM interaction (namely, left column $D=0.01$ and right column $D=0.5$).  P(s) on the top panels shows that  when the DM interaction DM interaction is small ($D=0.01<\epsilon_d$), the system is still nearly integrable and a distribution close to Poisson is obtained. Further increases of the DM strength causes the system to undergo a transition to the chaotic domain, the Wigner Dyson distribution is obtained. Further inspection can give the exact transition point called $D_c$. In the middle panels, we have depicted the corresponding of NPC values over the energy range using the site-basis. The regular system, left panel, shows a wide distribution over energy range, while the chaotic regime shows less fluctuations. Most eigenstates are concentrated in the middle of the spectrum where we also find the largest values of NPC. The bottom panels of Fig.~\ref{fig1}, show the eigenstate expectation values, $EEVs$, for the  magnetization on the defect site $\langle S_{d}^{z} \rangle$. The outcomes obtained for the delocalization measure, parallel  those for the $EEVs$. For small values of DM, (left panel), there are large fluctuations of the $EEVs$ of the $\langle S_{d}^{z} \rangle$ over the entire spectrum. One finds a very wide distribution of the values. As DM increases, and the system moves away from integrability, one can see that the fluctuations of $EEVs$ of nearby eigenstates reduce dramatically. A smooth behavior of the $EEVs$ with energy is achieved. This is what one needs for the validity of ETH. The comparison between the left and right columns make evident the strong connection between quantum chaos and thermalization. 
\par
  \subsection{XX model with the DM interaction}
We now investigate the effects f the DM interaction on the chain in the absence of the Ising interaction, $ J_{Z}=0$, with only XX coupling and a defect in the middle chain.
\par
\begin{figure}[h!]
\includegraphics[width=1.0\columnwidth]{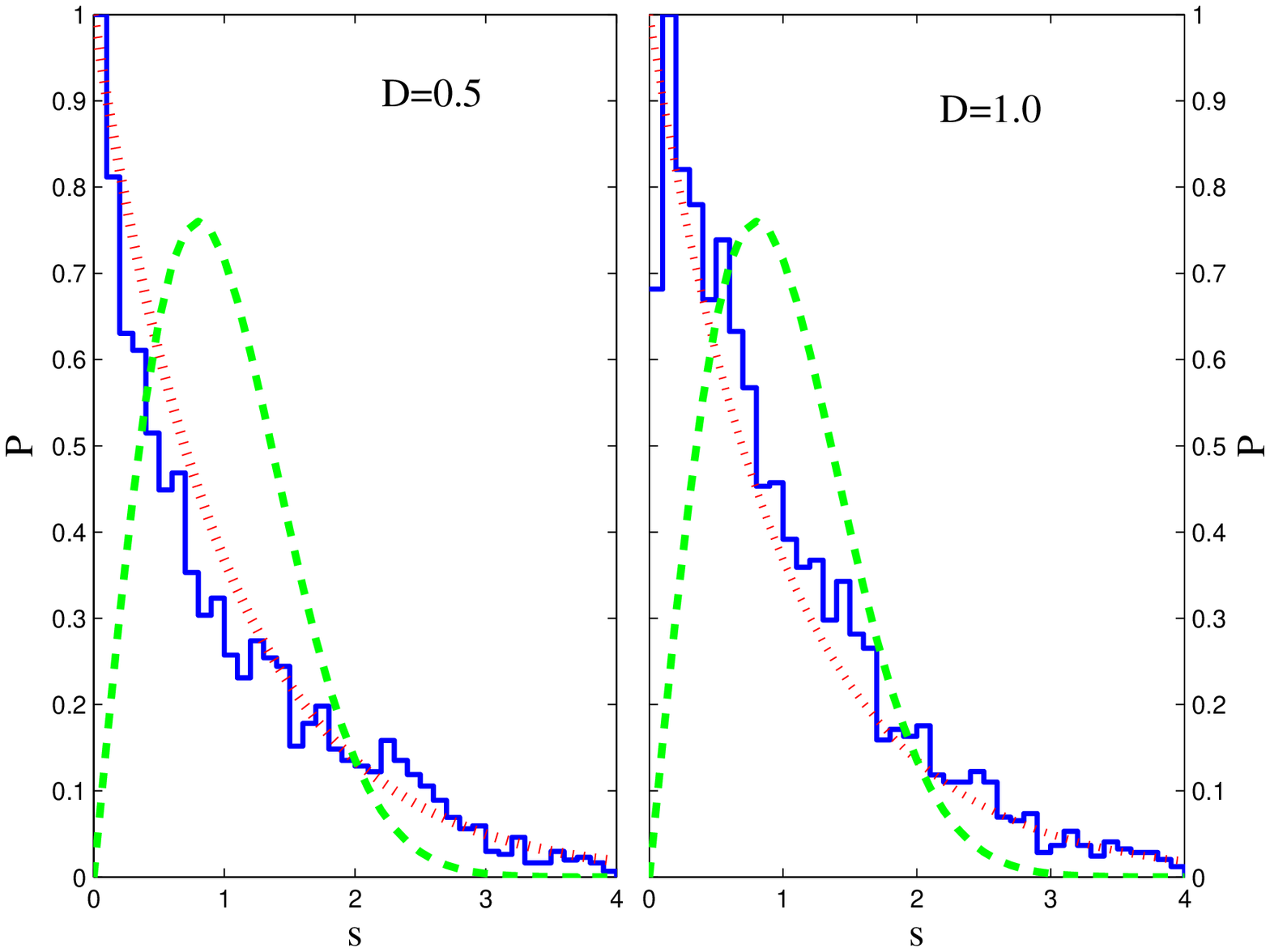}
\includegraphics[width=1.0\columnwidth]{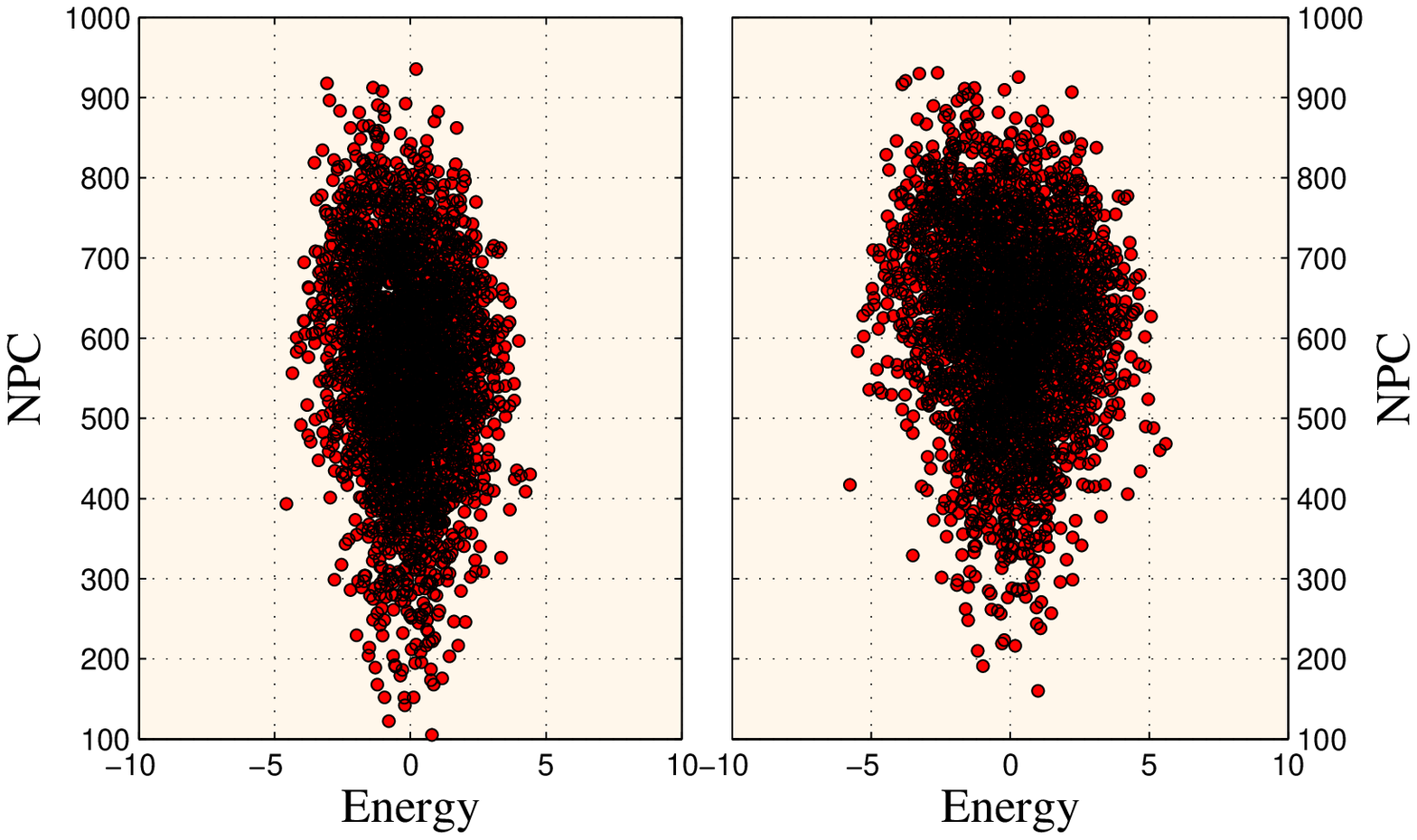}
\includegraphics[width=1.0\columnwidth]{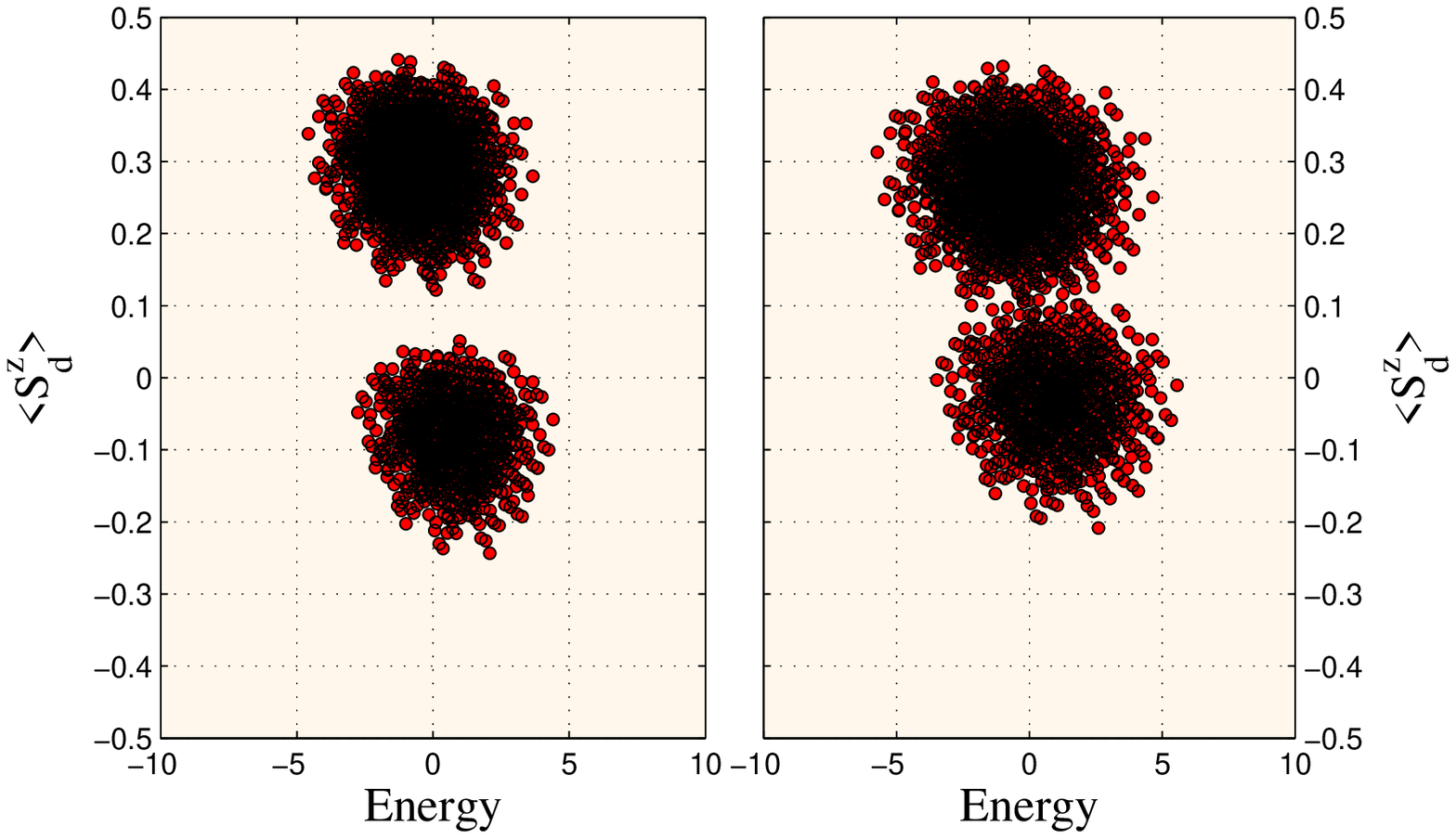}
\caption{(color online) Top, middle and bottom panels display the level spacing distribution $P(s)$, the NPC and $EEVs$ of $S_{d}^{z}$ vs energy for the full spectrum, respectively, for the Hamiltonian in Eq.~(\ref{Hamiltonian}) with $L=15$, $10$ up-spins and two values of DM interaction (left column $D=0. 5$ and right column $D=1. 0$). Other parameters are set to $J_{Z}=0$, $J=1$, $\epsilon=0.5$, $bin size=0.1$,  and a defect on site $d=7$. In both histogram level spacing distribution functions, top panels, dotted line gives the Poisson distribution and long-dashed line corresponds to the Wigner-Dyson distribution.}
\label{fig2}
\end{figure}
In Fig.~\ref{fig2}, we display  the level spacing distribution $P(s)$, NPC as a function of energy for all eigenstates, and $\langle S_{d}^{z}\rangle=\langle \psi_i | S_{d}^{z} | \psi_i \rangle$  for chain with length $L=15$ and for two values of the DM interaction (namely, left column $D=0.5$ and right column $D=1.0$). The top panels show the level spacing distribution $P(s)$. For both selected DM interaction strengths, the level spacing distribution is Poissonian. We have also explored other DM interaction strengths, ranging from $D=0.01 \cdot\cdot\cdot 1.0$, but we have not found any transition point from the integrable to the chaotic regime.  In the middle panels, we plot  the corresponding NPC distribution over the energy range using the site-basis.  As it can be seen, in both cases a large spread over the energy domain happens which could be a feature of a regular regime. The larger the DM interaction strength is, the more the spreading in energy becomes. 
\par 
In the chaotic regime, one expects less fluctuations in the values of NPC, indicating that the structures of the eigenstates with close energies are very similar. This is referred to as the uniformization of the eigenstates\cite{Percival73}, (note: the term uniformization refers to eigenstates that become so delocalized that they look alike and are similar random vectors.). However, by varying the value of the DM strength, the uniformization of the eigenstates did not happen. Large fluctuations is indeed what we expect from systems with Poisson distributions. Analogously to systems in the many-body localized phase, this system shows non-ergodic features and many-body eigenstates that are not thermal. Hence, initial states evolving under this Hamiltonian cannot relax to thermal equilibrium. 
 \par
  In the bottom panels, we illustrate the $EEVs$ of $S_{d}^{z}$ for all eigenstates of the Hamiltonian and for two values of DM interaction. We also checked other values of DM. In the absence of Ising interaction, for all values of DM, there are large fluctuations of the $EEVs$ of the observable over the entire spectrum which reflecting the results for NPC. Thus, ETH does not hold.

  \subsection{XXZ model with the DM interaction}
\begin{figure}[h!]
\includegraphics[width=1.0\columnwidth]{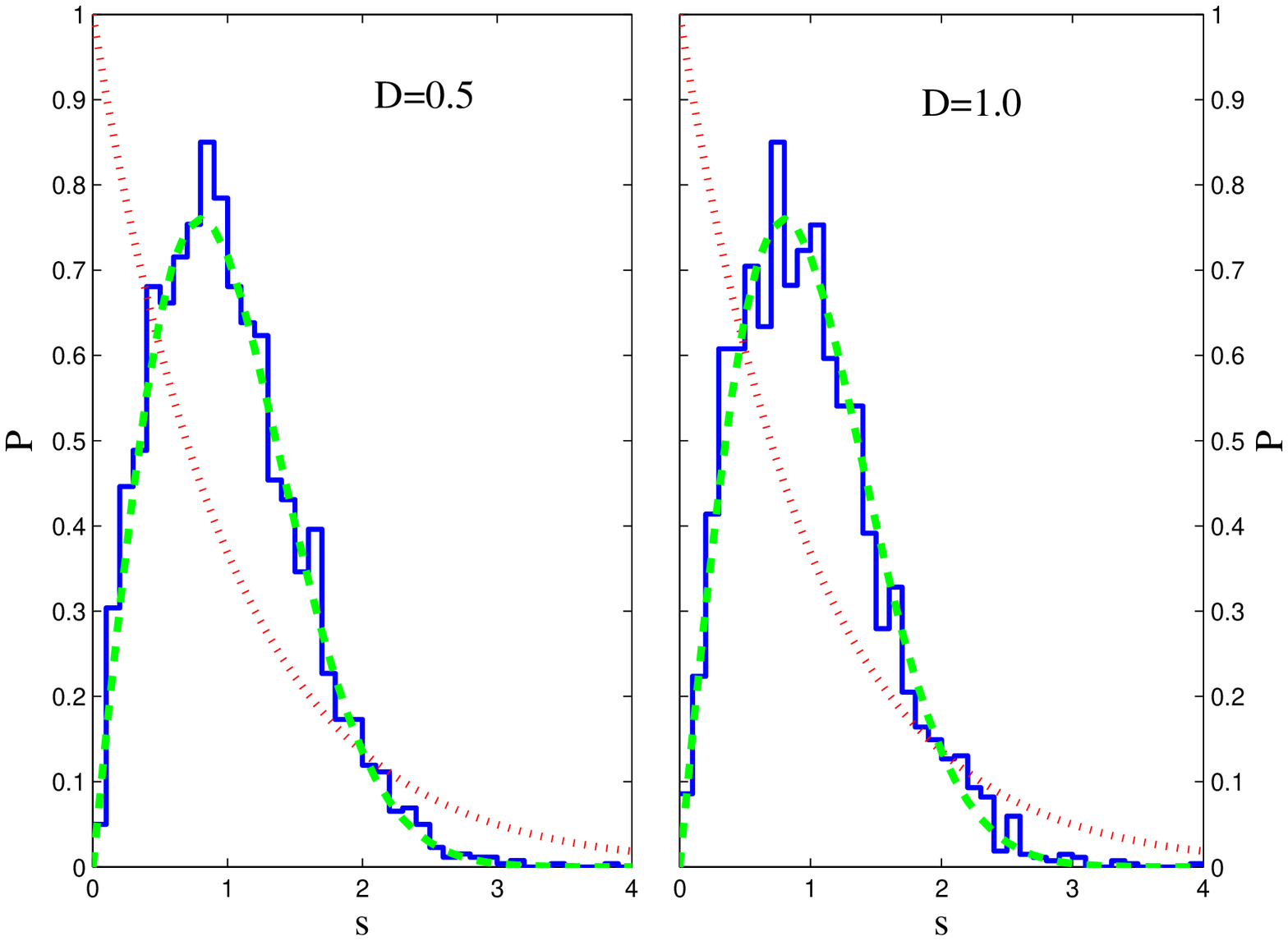}
\includegraphics[width=1.0\columnwidth]{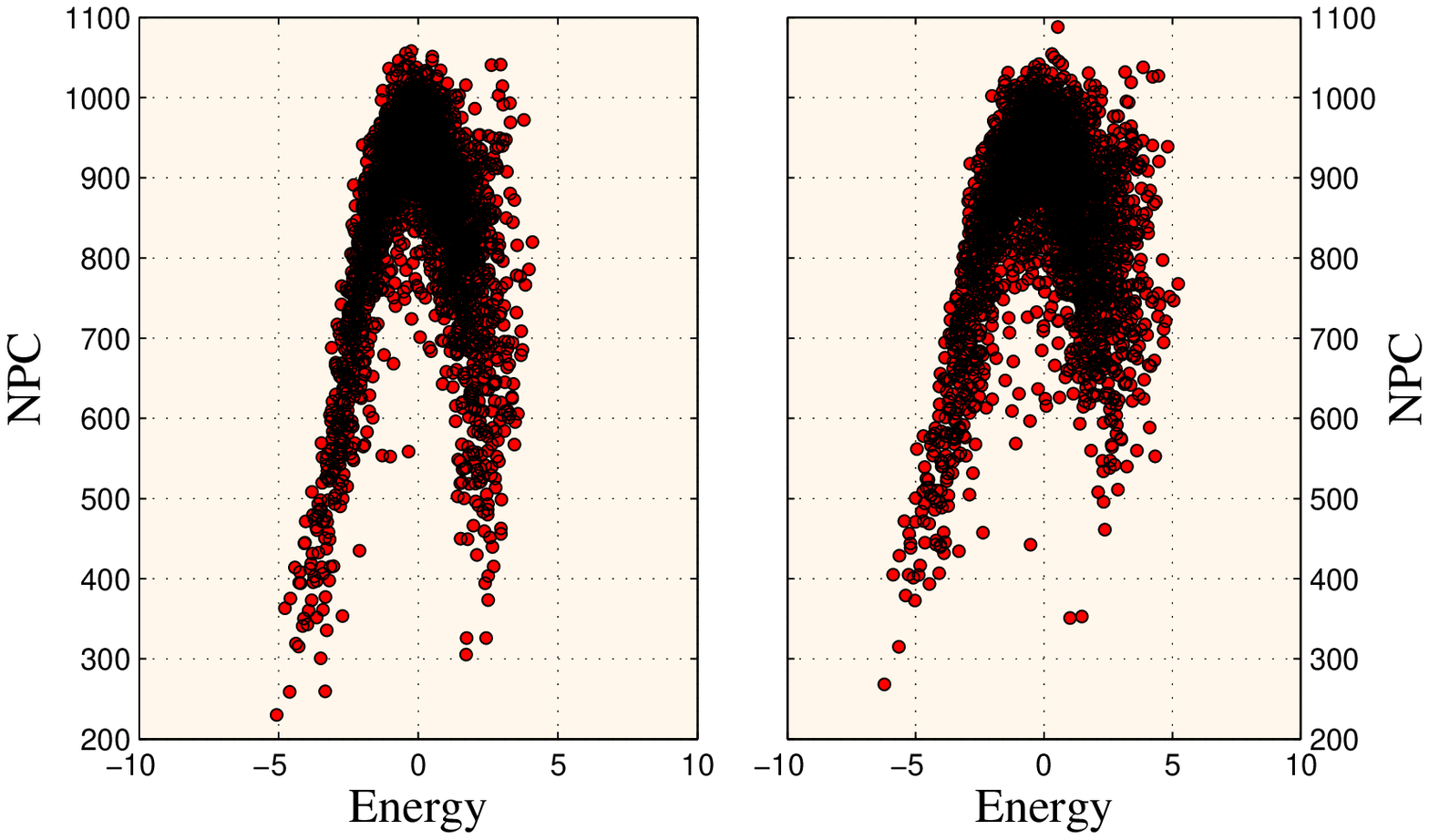}
\includegraphics[width=1.0\columnwidth]{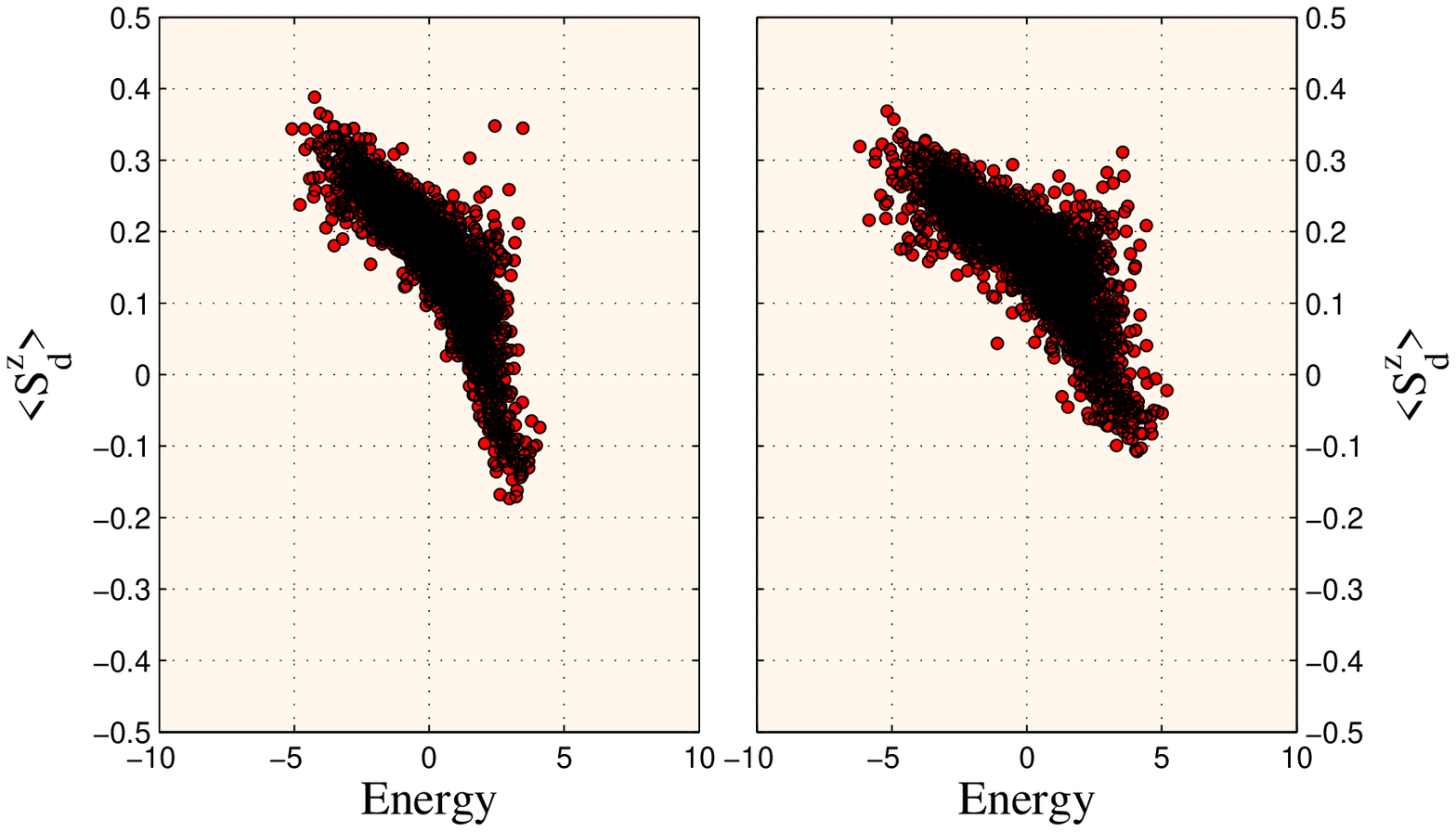}
\caption{(color online) Top, middle and bottom panels display the level spacing distribution $P(s)$, the  NPC and $EEVs$ of $S_{d}^{z}$ vs energy for the full spectrum, respectively, for the Hamiltonian in Eq.~(\ref{Hamiltonian}) with $L=15$, $10$ up-spins and some of the DM interaction values (left column $D=0. 5$ and right column $D=1. 0$). Other parameters are set to $J_{Z}=0.5$, $J=1$, $\epsilon=0.5$, $bin size=0.1$,  and a defect on site $d=7$. In both histogram level spacing distribution function, top panels, dotted line gives the Poisson distribution and long-dashed line corresponds to the Wigner-Dyson distribution}
\label{fig3}
\end{figure}

Here, we analyze the effect of the DM interaction on the anisotropic spin-1/2 XXZ chain in the presence of an external magnetic field, $h\neq0$, and a defect in the middle of the chain. In the case of $J>J_{Z}$(gapless phase), we computed numerically the level spacing distribution $P(s)$, the NPC distribution and $\langle S_{d}^{z}\rangle=\langle \psi_i | S_{d}^{z} | \psi_i \rangle$ for some values of the anisotropy $D$. Results are depicted in Fig.~\ref{fig3}. For all the DM interaction strength that we consider, the system lives in the chaotic regime, as can be seen from the Wigner-Dyson distributions in the top row of the Fig.~\ref{fig3}. In the middle panels, we show  the corresponding NPC distribution over the energy range using the site-basis. The panels show, in both cases, that the NPC is a smooth function of energy indicating the uniformization of the eigenstates, specially in the middle of the energy spectrum, which is a signature of the chaotic regime. Therefore, the XXZ model with added DM interaction shows ergodic features including thermal eigenstates. Generic initial states evolving according to this Hamiltonian should relax to thermal equilibrium. 
\par
In the bottom row of the Fig.~\ref{fig3}, we illustrate the $EEVs$ of $S_{d}^{z}$ for all eigenstates of the Hamiltonian and for two values of DM interaction. We  also checked other values of DM and observed that for all values of the DM strength that are not much larger than the parameters $J$, $J_z$, and $\epsilon_d$, these fluctuations reduce in the center of the spectrum and a smooth behavior of the $EEVs$ with energy is achieved indicating that ETH is valid.
\begin{figure}
\includegraphics[width=1.0\columnwidth]{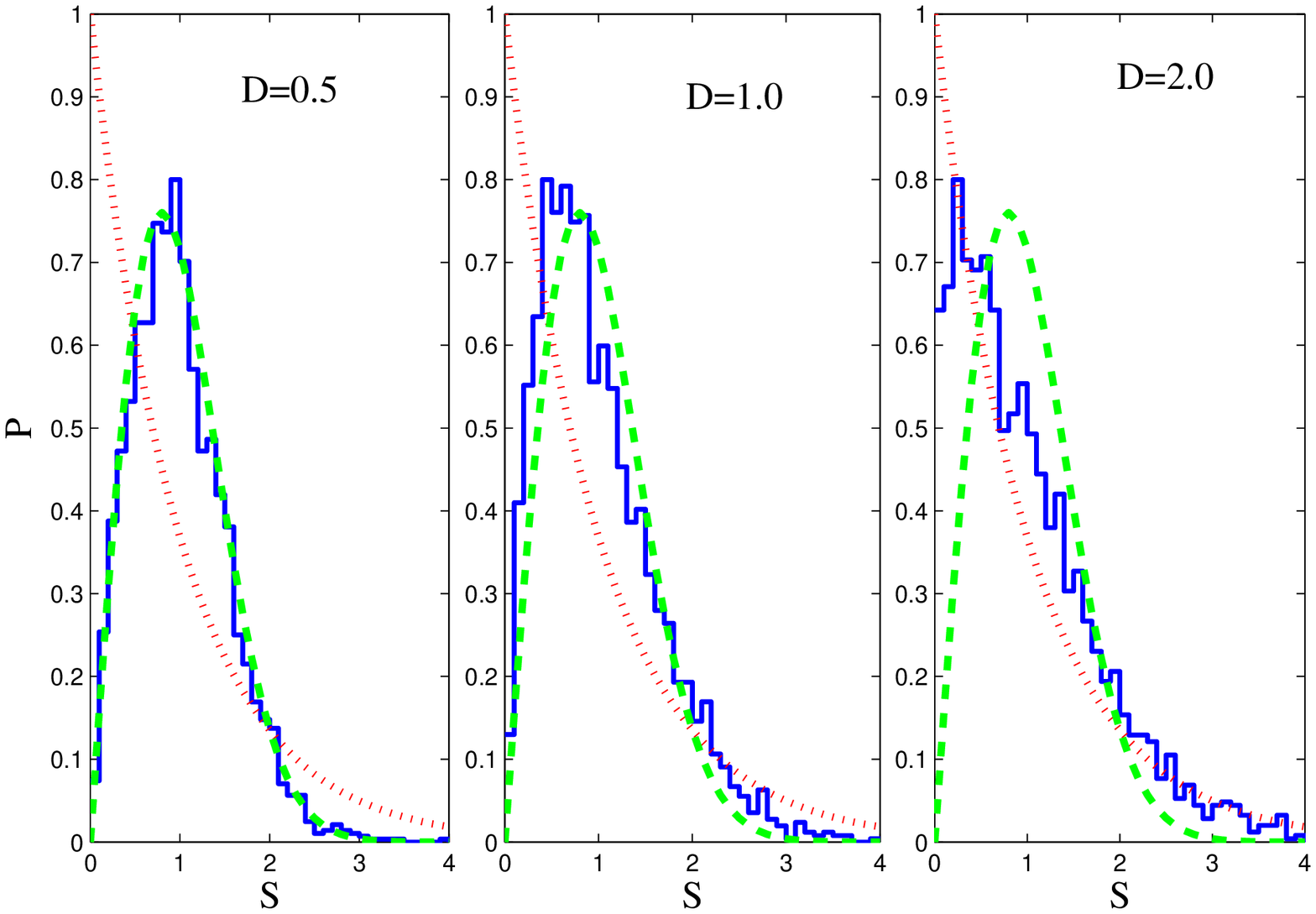}\\
\includegraphics[width=1.0\columnwidth]{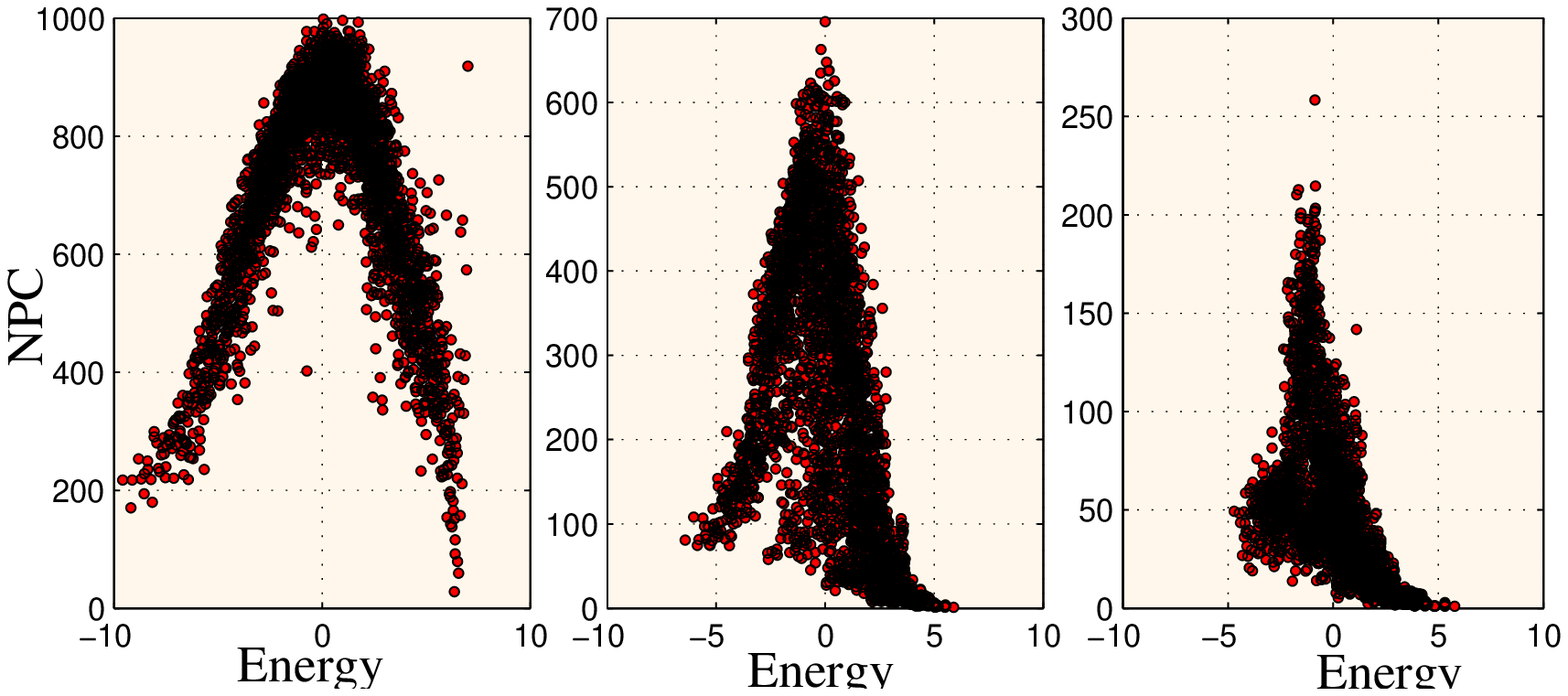}\\
\includegraphics[width=1.0\columnwidth]{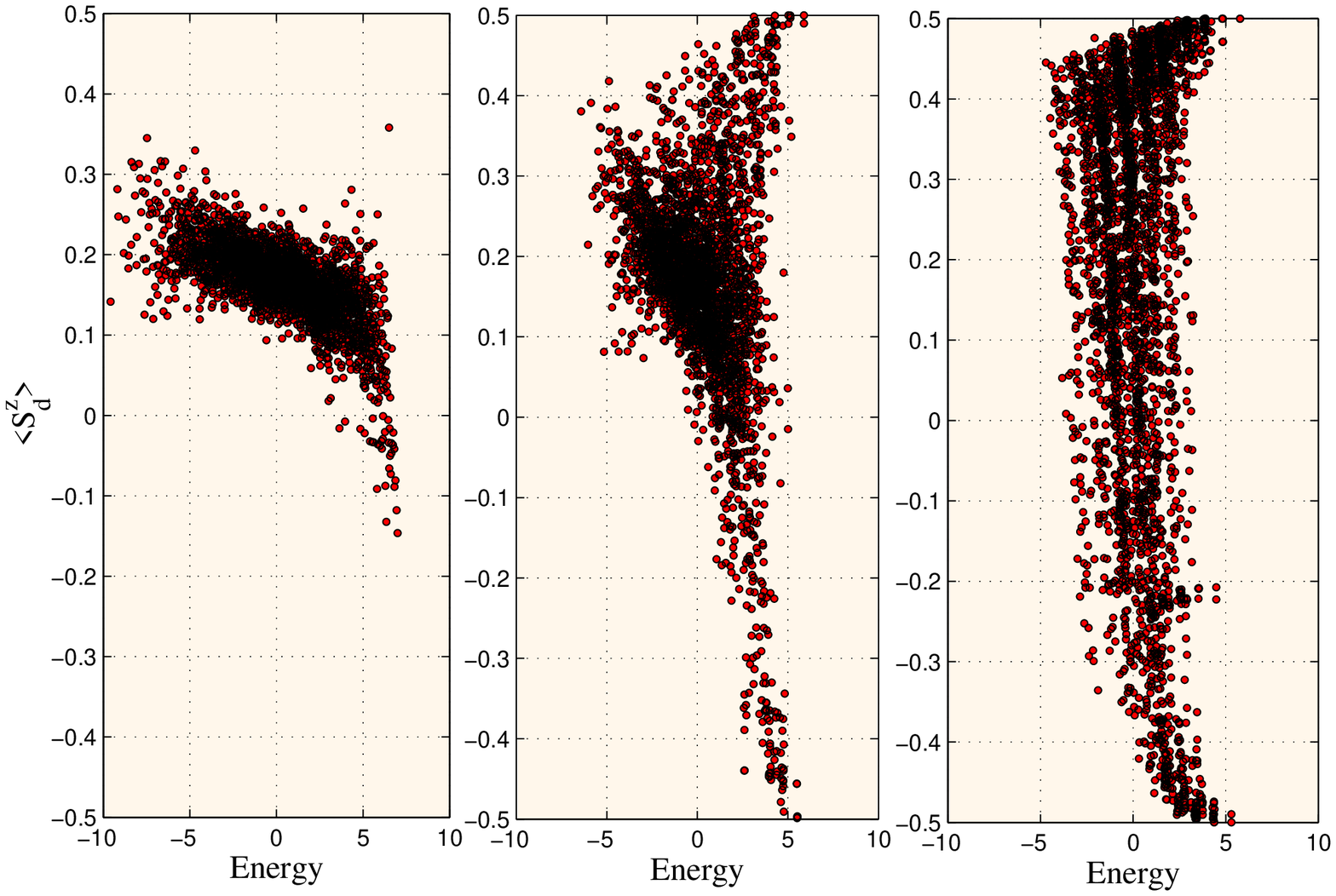}
\caption{(Color online)Top, middle and bottom panels are depicting the level spacing distribution $P(s)$, the NPC and $EEVs$ of $S_{d}^{z}$ vs energy for the full spectrum, respectively). Other parameters are set as $J=0.1$, $J_{Z}=2.0$, $\epsilon=0.5$, $bin size=0.1$, $10$ up-spins and a defect on site $d=7$. In histograms (top row) dotted line gives the Poisson distribution and long-dashed line corresponds to the Wigner-Dyson distribution.}
\label{fig4}
\end{figure}
\par
In the case of $J_{z}\gg J$, the XXZ anisotropic spin-1/2 chain in the absence of the DM interaction and with only nearest-neighbor interaction, even with the presence of a defect in the middle, is nearly integrable, since we approach the limit of the Ising model. In this case, the level spacing distribution is close to Poisson (not shown here). When we add the DM interaction with different values in the chain, depending on the DM value, the system can be chaotic and have a Wigner-Dyson distribution, as can be seen from Fig.~\ref{fig4}. By tuning the DM interaction, the system gradually tends to show chaotic characteristics which are visible in the case $D=0.5$ (see the top row of Fig.~\ref{fig4}).  By further increasing the DM value, $D=1.0, 2.0$, the system undergoing a dynamical transition and moving away from the integrable domain, in result the Poisson distribution is obtained. 
\par
In the bottom panels, we depict the $EEVs$ of $S_{d}^{z}$ for all eigenstates of the Hamiltonian and for different values of DM interaction. For small values of DM for example $D=0.5$, (left panel), there are less fluctuations of the $EEVs$ of the observable over the entire spectrum, smooth behavior of $EEVs$ with energy is achieved, and ETH becomes valid. Increasing gradually the DM interaction develops a transition from chaotic to integrable regime,  these is visible from large fluctuations of the $EEVs$ of $S_{d}^{z}$, in result ETH does not happen. 
\section{Conclusion}\label{sec4}
Using  exact diagonalization, the Heisenberg spin chain with  a local defect placed in the middle of the chain was studied in the context of the quantum chaos. The main concern of  this work was to explore the emergence of quantum chaos in the presence of the spin-orbit interaction as a pure quantum interaction.  To this end, we have considered the Ising, the XX and the XXZ limits of the Heisenberg chain with added Dzyaloshinskii-Moriya(DM) interaction. Our results show that quantum chaos develops in the XXZ and Ising chains, but not in the XX model. In the case of the XXZ model, we also investigated the interplay between the anisotropy and the DM interaction strength. Quantum chaos is the cause of thermalization in isolated many-body quantum systems. We showed that in the chaotic limit, the structures of the eigenstates with close energies are very similar, resulting in small fluctuations of the eigenstate expectation values of the magnetization. This indicates that the ETH should be valid in the chaotic domain. We believe that our finding may sound interesting for those are active in this line.
\section{acknowledgments}
J. Vahedi acknowledge  the Center for Theoretical Physics of Complex Systems Institute for Basic Science, Daejeon, Korea,  for their warm hospitality and financial support during his visit where part of this work was done.
\vspace{0.3cm}

\end{document}